\documentclass[letter]{aa}  

\usepackage{graphicx}
\usepackage{txfonts}
\defcitealias{Mejia-NarvaezBruzual2021}{MN21}
\defcitealias{Bland-HawthornGerhard2016}{BH16}

\begin{document}

   \title{Spectral Evidence of Solar Neighborhood Analogs in CALIFA Galaxies}

   \subtitle{}

   \author{
        A. Mej\'ia-Narv\'aez\inst{1}
        \and
        S.~F. S\'anchez\inst{1}
        \and
        L. Carigi\inst{1}
        \and
        J.~K. Barrera-Ballesteros\inst{1}
        \and
        N. Drory\inst{2}
        \and
        C. Espinosa-Ponce\inst{1}
    }

   \institute{
        Instituto de Astronom\'ia, Universidad Nacional Aut\'onoma de M\'exico, A. P. 70-264, C.P. 04510, M\'exico, D.F., M\'exico
        \and
        McDonald Observatory, The University of Texas at Austin, 1 University Station, Austin, TX 78712, USA
    }

   \date{Received September 15, 1996; accepted March 16, 1997}

 
  \abstract
   {}
   {
   We introduce a novel non-parametric method to find solar neighborhood analogs (SNAs) in extragalactic IFS surveys. The main ansatz is that the physical properties of the solar neighborhood (SN) should be encoded in its optical stellar spectrum.
   }
   {
   We assume that our best estimate of such spectrum is the one extracted from the analysis performed by the Code for Stellar properties Heuristic Assignment (\textsc{CoSHA}) from the MaStar stellar library.
    It follows that finding SNAs in other galaxies consist in matching, in a $\chi^2$ sense, the SN reference spectrum across the optical extent of the observed galaxies. We apply this procedure to a selection of CALIFA galaxies, by requiring a close to face-on projection, relative isolation and non-AGN. We explore how the local and global properties of the SNAs (stellar age, metallicity, dust extinction, mass-to-light ratio, stellar surface mass and star-formation densities and galactocentric distance) and their corresponding host galaxies (morphological type, total stellar mass, star-formation rate, effective radius) compare with those of the SN and the Milky Way (MW).
   }
   {
   We find that SNAs are located preferentially in S(B)a~--~S(B)c galaxies, in a ring-like structure, which radii seems to scale with the galaxy size. Despite the known sources of systematics and errors, {most} properties present a considerable agreement with the literature on the SN. We conclude that the solar neighborhood is relatively common in our sample of SNAs.
    Our results warrants a systematic exploration of correlations among the physical properties of the SNAs and their host galaxies. We reckon that our method should inform current models of the galactic habitable zone in our MW and other galaxies.
   }
   {}


   \maketitle
%

\section{Introduction}

\label{sec:intro}

Studies of the solar vicinity, whether such vicinity involves only a few hundreds of stars around the Sun or the whole Milky Way (MW), have the potential to answer fundamental questions about the formation of our galaxy and its satellites \citep[e.~g.,][]{HelmiBabusiaux2018, Ruiz-LaraGallart2020}, the formation of the Sun and its planets \citep[e.~g.,][]{RaymondIzidoro2020} and ultimately the physical conditions needed to sustain life as we know it \citep[e.~g.,][]{GonzalezBrownlee2001}. In an attempt to characterize the most likely zone in our MW with the conditions needed for life to sprout and thrive, \citet{GonzalezBrownlee2001} introduced the Galactic habitable zone (GHZ), a set of values for relevant physical properties that should constrain the habitability of different locations in the MW and in other galaxies. They suggested the GHZ as a ring-like structure located in the thin disk that extends with time toward the outskirts of the disk.

{The GHZ was latter explored by \citet{LineweaverFenner2004}. 
Based on chemical evolution models, those authors argued that the width of the ring should grow as a function of the cosmic time, due to the inside-out evolution of the disk (from $6\,$ to $10\,$kpc). Similar results were found by \citet[][$7$~--~$9\,$kpc]{SpitoniMatteucci2014}.
By considering the stellar density, \citet{Prantzos2008} found that the most probable GHZ is in the inner disk ($\sim3\,$kpc), later confirmed by \citet{GowanlockPatton2011}. In the cosmological context, the GHZ has also been characterized through simulations, with the predictions that our Sun is not in the most probable GHZ, instead being either in the lower limit of the galactocentric distance distribution \citep[e.~g.,][]{VukoticSteinhauser2016} or in the upper limit \citep[e.~g.,][]{ForganDayal2017} and, more interestingly, not in the most likely host galaxy type \citep[e.~g.,][]{ZackrissonCalissendorff2016, GobatHong2016}. Such theoretical results lack so far of a solid observational support.}


The physical properties of the MW and the solar neighborhood (SN) have been widely studied in different surveys e.~g., SEGUE \citep{YannyRockosi2009} APOGEE \citep{HoltzmanShetrone2015}, GALAH \citep{DeSilvaFreeman2015} and Gaia \citep{GaiaCollaborationBrown2018}. As a matter of fact, we have drawn a clear picture of our MW \citep[see][for a review; \citetalias{Bland-HawthornGerhard2016} hereafter]{Bland-HawthornGerhard2016} and the SN \citep[e.~g.,][]{BuderLind2019}, a knowledge that has been fed to simulations to reproduce observations with success \citep[e.~g.,][]{PrantzosAbia2018}. At global (i.~e., the typical scale of a galaxy) and extragalactic scales, we have sampled hundreds of MW analogs in large surveys \citep[e.~g.,][]{Fraser-McKelvieMerrifield2019, BoardmanZasowski2020}, leading to the conclusion that the MW is common among its analogs, but relatively rare among a complete sample of galaxies of the local Universe. At local scales (i.~e., kpc scales and below), however, we have still to make progress in finding SN analogs (SNAs) in extragalactic surveys.

In this letter we seek to bridge the gap between simulations of the evolution of the SN and the whole MW, and current observations of its physical properties. Therefore, we use a non-parametric method to find SNAs in other galaxies that have been observed using Integral Field Unit (IFU). The main premise in our endeavor is that the physics of our SN is encoded in its integrated optical spectrum. Since we know for a fact that in our SN life exists, if follows that the physical conditions defining our portion of the GHZ should also be encoded in such spectrum. We search for answers to the following questions: are SNAs properties, as found in other galaxies, similar to our own SN? will this rationale lead to the SN being a typical or an atypical environment in galaxies? in what type of galaxies -- morphologically speaking -- are SNAs more likely to be found? In particular we look for a characterization of the galactocentric distance distribution of the SNAs and to know if it scales with the host galaxy size. A natural follow up is to know if the Sun is in an expected location or if it is an outlier in this distribution. This letter is organized as follows: in \S~\ref{sec:sample} we describe the samples and analysis methods; in \S~\ref{sec:results} we present our main findings; and finally we discuss and conclude in \S~\ref{sec:conclusions}.

\section{Data and analysis}\label{sec:sample}


\subsection{MaStar as a spectroscopic sample of solar neighborhood stars}\label{sec:mastar}

MaStar \citep{YanChen2019} is a stellar library for the MaNGA survey \citep{BundyBershady2015, DroryMacDonald2015}. \citet[][hereafter \citetalias{Mejia-NarvaezBruzual2021}]{Mejia-NarvaezBruzual2021} labeled $\sim22\,$k unique stars in this library using \textsc{CoSHA}, a heuristic Machine Learning approach. One by-product of the analysis in \citetalias{Mejia-NarvaezBruzual2021} that is relevant to the present letter is the partial volume correction implemented using the color-magnitude diagram (CMD) sampled by Gaia DR2 \citep{GaiaCollaborationBrown2018}. The Gaia survey has been implemented in the recent past to analyse the solar neighborhood properties \citep[e.~g.,][]{DingZhu2019, Sollima2019, GontcharovMosenkov2021, AlzateBruzual2021}, therefore it is a good survey to draw a photometric sample \citep[e.~g.,][]{EvansRiello2018} of the solar surroundings. Knowing this, \citetalias{Mejia-NarvaezBruzual2021} calculated a partial volume correction using the relation:
\begin{equation}\label{eq:vcor}
V_\mathrm{cor} = \frac{\mathrm{CMD}_\mathrm{MaStar}}{\mathrm{CMD}_\mathrm{Gaia}},
\end{equation}
where $\mathrm{CMD}_\mathrm{MaStar}$ and $\mathrm{CMD}_\mathrm{Gaia}$ represent the PDF distribution of stars in the extinction corrected CMD sampled by MaStar and Gaia \citep{GaiaCollaborationBabusiaux2018}, respectively \citep[e.~g.,][]{WallJenkins2003, Rodriguez-PueblaPrimack2017, SanchezAvila-Reese2019}. The ``partial'' character of this volume correction comes from the fact that Gaia is not a complete sample of the stars in the solar neighborhood. However it is, to our best knowledge, the most complete sample to date. The fact that the volume-corrected distributions of MaStar chemical abundances ($\left[\mathrm{Fe}/\mathrm{H}\right]$ and $\left[\alpha/\mathrm{Fe}\right]$) resemble those from independent studies in the SN \citepalias[c.~f. Fig. 9 in][]{Mejia-NarvaezBruzual2021}, is encouraging. Hence, we can assert that \emph{the weighted averaged MaStar spectrum, with weights $V_\mathrm{cor}$ is, by design, our best estimate for the solar neighborhood optical spectrum.} In the following, we allude to this spectrum, $f_\lambda^\mathrm{SN}$, as our spectroscopic reference of the solar neighborhood or simply the SN spectrum.

\subsection{A solar neighborhood definition}\label{sec:solar-neighborhood}

The solar neighborhood is usually defined as the volume enclosed within a set radius around the Sun. There is no consensus, however, on the value of such radius and the literature on MW studies spans ranges from a few tens of pc to $1\,$kpc, depending on the specific subject of the research \citep[e.~g.,][]{VergelyFerrero1998, AniyanFreeman2016}. Here, we define the radial scale of the solar neighborhood \emph{a posteriori}, as the standard deviation of the volume corrected distance distribution of stars in the MaStar sample: $r_\mathrm{SN}=1.24\,$kpc. This definition is consistent with the upper limit quoted above. It is also consistent with the typical physical resolution in the CALIFA sample, at the average redshift $\sim0.8\,$kpc ($\textrm{FWHM}=0.3$~--~$1.8\,$kpc), which is indeed a convenient coincidence for the current study (although it does not impose a strong limitation). We note, that this definition is purely observational and that we may have a contribution from halo stars. However, as we will see below, any potential bias due to this selection should be accounted for by the projection effects of CALIFA galaxies.

\subsection{A golden sample from CALIFA}\label{sec:califa}

CALIFA \citep{SanchezKennicutt2012} is an integral field spectroscopic (IFS) survey of $\sim1000$ galaxies in the nearby universe $z\sim0.015$, complete to $M_\star\sim10^{11.4}\,$M$_\sun$ and spanning all morphological types \citep{WalcherWisotzki2014, GalbanyAnderson2018}. We note that CALIFA is the only IFS survey that covers up to $2.5\,r_\textrm{eff}$, i.~e., most of the optical extension of galaxies is within the field-of-view (FoV) of the instrument, which is a relevant trait in our study \citep[c.~f. Table 1 in][]{Sanchez2020}. The physical properties of these galaxies have been estimated and summarized using the \texttt{Pipe3D} pipeline \citep{SanchezPerez2016a, SanchezPerez2016b}, which data-products have been extensively used in numerous publications \citep[e.~g.,][]{Lopez-CobaSanchez2020, Mejia-NarvaezSanchez2020, ValerdiBarrera-Ballesteros2021, Barrera-BallesterosHeckman2021, Espinosa-PonceSanchez2022}. We focus our attention in galaxies that are not highly inclined in order to locate SNAs across their optical extent. In particular, we have three requirements for a galaxy to belong to our golden sample: \textit{(i)} the inclination angle is $\leq60\,$deg, also avoiding known issues \citep[e.~g.,][]{Ibarra-MedelAvila-Reese2019}, \textit{(ii)} we have a reliable morphological classification; \textit{(iii)} the galaxy has no evidence of nearby companions, ongoing collisions or post-merger signatures;\footnote{We will analyze interacting galaxies in a another paper.} and \textit{(iv)} the galaxy most not host an AGN according to \citet{LacerdaSanchez2020}. After applying these constraints our initial sample is reduced to $330$ galaxies, ($\sim1/3$) of the original CALIFA sample. This golden selection poses the possibility of finding SNAs in galaxies that hold no \emph{a priori} resemblance with the MW.

\subsection{Finding solar neighborhood analogs}\label{sec:finding-sna}

We look for the SNA across the optical extent of CALIFA galaxies by exploring the spatial distribution of the ${\chi^2}$, in which the observed spectra in each spaxel is compared with our reference spectrum for the SN in the MW (\S~\ref{sec:solar-neighborhood}).
For this purpose, the SN reference spectrum is convolved with a Gaussian function and dust attenuated using the \citet{CardelliClayton1989} model, following the procedures in \texttt{pyFIT3D} \citep[][Eq.~1]{LacerdaSanchez2022}. Hence, we account for (1) the redshift and velocity rotation of the galaxy, (2) the line-of-sight velocity dispersion, and (3) the differential extinction due to projection and line-of-sight effects. To further account for the differential physical spatial resolution of each spaxel ($\sim1\,\arcsec$) at the observed galaxy redshift, the CALIFA cubes are convolved with a Gaussian kernel to match the resolution of our SN region ($\sim1\,$kpc, \S~\ref{sec:solar-neighborhood}).

Based on the ${\chi^2}$-matching we perform for each spaxel,
we label as SNA those regions in which the ${\chi^2}_{ij}$ is within the range $0.7$~--~$1.3$, avoiding at the same time, the spaxels that mismatch with our reference spectrum ($>1.3$) and those with over estimated uncertainties ($<0.7$).

\section{Results}\label{sec:results}

\begin{figure*}[ht!]
\includegraphics[width=\textwidth]{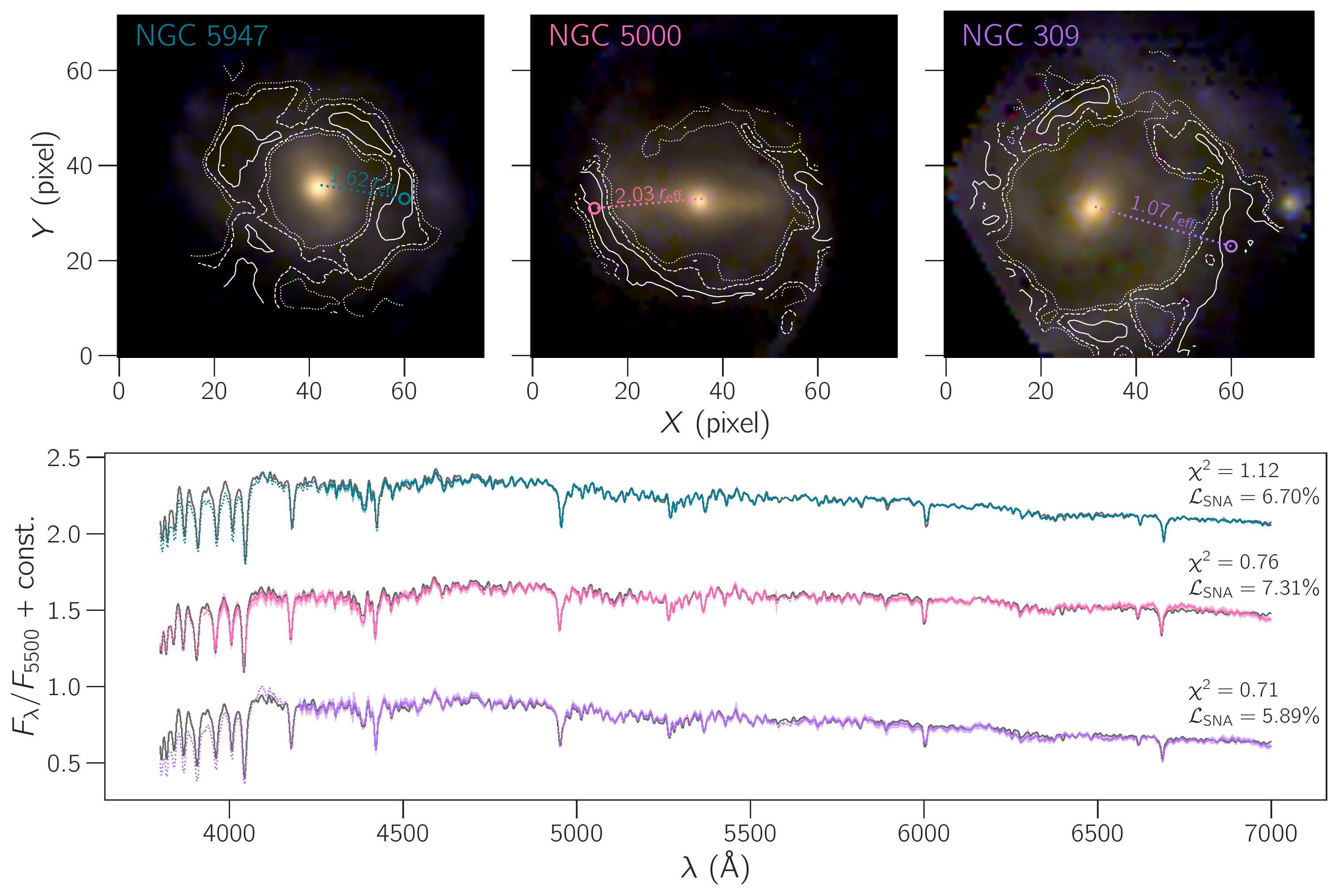}
\caption{\textit{Top panels:} The RGB (R: $6450\,$\AA{}, G: $5375\,$\AA{} and B: $3835\,$\AA{}) composite images of galaxies NGC~$5947$, NGC~$5000$ and NGC~$309$ in our golden sample described in \S~\ref{sec:califa} that host likely SNA regions. {The likelihood map is overlaid in each case with contours enclosing} the $1\sigma$ (solid), $2\sigma$ (dashed) and $3\sigma$ (dotted), respectively. {The location of the maximum likelihood in each case (colored circle) and its galactocentric distance (dashed colored line) are shown.} \textit{Bottom panel:} the spectrum corresponding to the maximum likelihood spaxel (solid line in color) along with the propagated ($1\sigma$) error spectrum is also shown as a shaded region. We highlight the regions where the error spectrum was masked out (dotted). The SNA spectrum (gray) calculated as described in \S~\ref{sec:finding-sna} is also shown. The resulting $\chi^2$ and the likelihood integrated within $r_\mathrm{pix}$ around the maximum are also shown.
\label{fig:sna-showcase}}
\end{figure*}

Fig.~\ref{fig:sna-showcase} summarizes our main results for three galaxies hosting SNA regions, as an illustrative example: NGC~$5947$, NGC~$5000$ and NGC~$309$. For each galaxy we show the contour of the likelihood map ($\mathcal{L}_{ij}\propto\exp{\left[-{\chi^2}_{ij}/2\right]}$) on top of the RGB composite image, highlighting the location of the most likely SNA. In addition we show the corresponding spectra of those regions together with the reference SNA spectrum.
{We observe that} the SNA spectrum of the three example galaxies is well within the uncertainty band, $\sigma_\lambda$ (shaded region), for most of the wavelength range. A higher $\mathcal{L}_\textrm{SNA}$ indicates that the likelihood of finding a SNA within the target galaxy is more concentrated around a specific region. In top panels of Fig.~\ref{fig:sna-showcase}, NGC~$5000$ shows the most concentrated likelihood of hosting SNA spaxels, with an apparent azimuthal symmetry. At first sight, this pattern is also followed by NGC~$5947$ and NGC~$309$, whereby likely SNA spaxels group in a ring-like structure. Interestingly, the radius of such structure seems to depend on the galaxy's apparent size. We will explore this below.

Repeating the above analysis for each galaxy we find SNAs in only $61$ out of $330$ galaxies ($\sim20\%$). Table \ref{tab:sna-chars} shows the physical properties of the MW and in our solar vicinity (column 2) together with those of the host galaxies and the corresponding most likely SNA regions (column 3). A detailed exploration of these properties will be presented in a forthcoming study. We present here a brief summary of such exploration.

\begin{figure}
\includegraphics[width=\columnwidth]{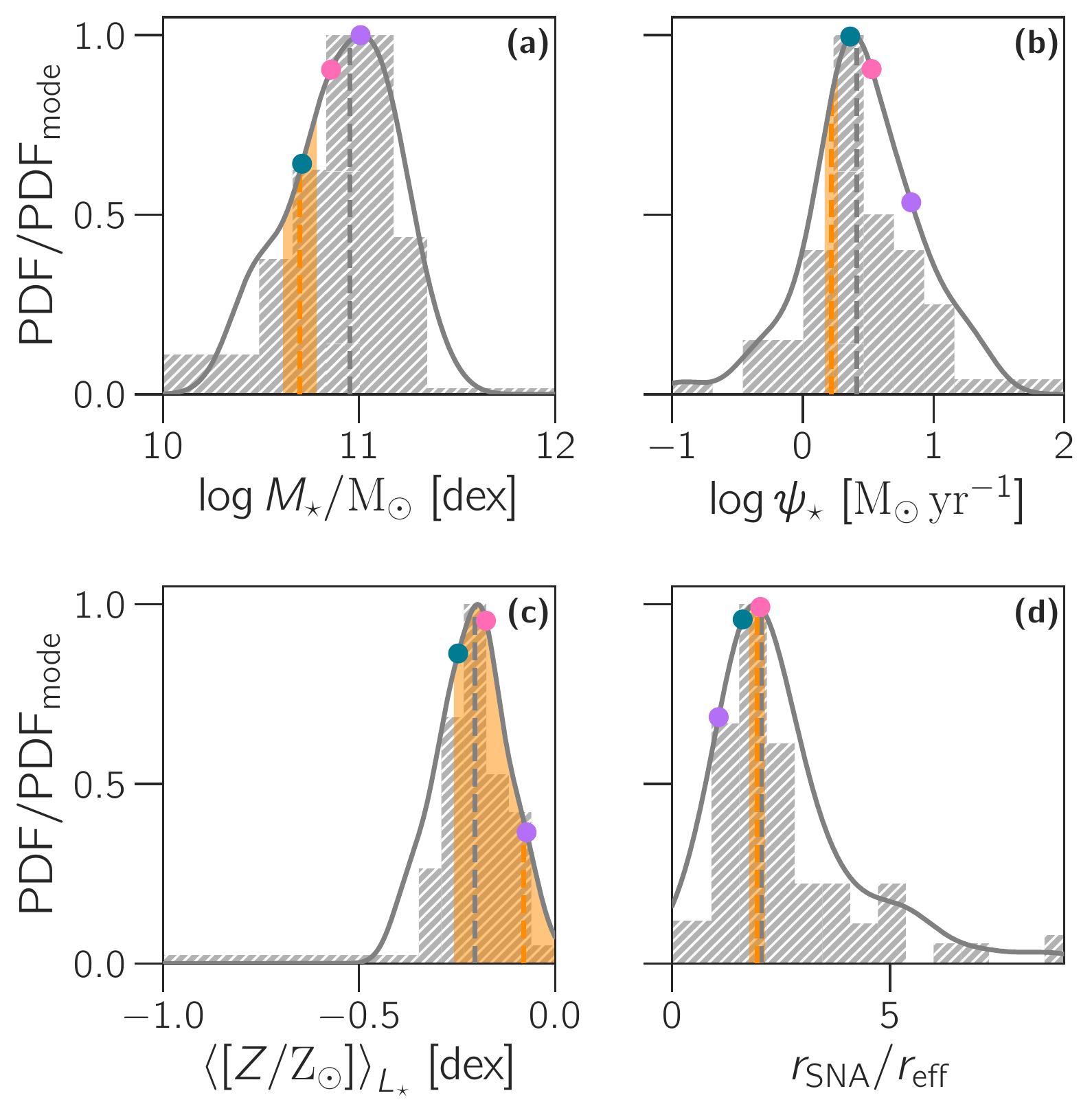}
\caption{{Distributions of global \textit{(a)} stellar mass and \textit{(b)} SFR; and local \textit{(c)} luminosity-weighted stellar metallicity and \textit{(d)} galactocentric distances normalized by the host effective radius, for the $61$ host and SNAs (gray shaded histogram and density plot), respectively. The corresponding median of each distribution is shown (gray dashed line). The $3$ example galaxies from Fig.~\ref{fig:sna-showcase} are also highlighted (circles). The vertical orange region indicates the galactocentric distance of the Sun with respect to the $r_\textrm{eff}$ of the MW ($\pm1\sigma$) around the best estimate (orange dashed line)}.
\label{fig:sna-pars}}
\end{figure}

Despite remaining agnostic about the global and local physical properties of potential host galaxies, we find that \emph{most} likely SNA regions (within $1\sigma$) are more frequent in spiral galaxies ranging Sa~--~Sc and that these regions seem to lie preferentially in between arms. As a matter of fact, this is the expectation in the MW (Sb~--~Sbc), where the Sun is known to be located between Perseus and Carina-Sagittarius arms \citepalias[e.~g.,][]{Bland-HawthornGerhard2016}. The {median} global stellar mass {(Fig.~\ref{fig:sna-pars}a)}, $\log{M_\star/\textrm{M}_\sun}\sim10.91\pm0.26$ and star-formation rate {(SFR; Fig.~\ref{fig:sna-pars}b)}, $\log{\psi_\star}\sim0.46\pm0.43\,$M$_\sun\,$yr$^{-1}$, for our sample of host galaxies, is also consistent with reported values for the MW \citep[e.~g.,][]{Fraser-McKelvieMerrifield2019}. However, we note that the corresponding MW values are within $1\sigma$ of the host galaxies sample, closer to its lower limit (c.~f., Table~\ref{tab:sna-chars}). 
We know that most global properties of galaxies scale with its size \citep[e.~g.,][and references therein]{SanchezWalcher2021}, therefore we also compare the median effective radius of the host galaxies sample, $r_\textrm{eff}\sim5.76\pm3.32\,$kpc to the Galaxy value. Given the location of the Sun within the MW disk, a precise measure of its radial scale-length is difficult \citepalias[see,][for a review]{Bland-HawthornGerhard2016}. We assume the reviewed radial scale-length $\sim2.5\pm0.4\,$kpc. To convert to effective radius we multiply by $1.68$ \citep{SanchezRosales-Ortega2014},
resulting in an estimated effective radius for the MW of $r_\textrm{eff}\sim4.2\pm0.4\,$kpc. As in the case of the stellar mass and the SFR, the MW lies within $1\sigma$ of the distribution of SNA host galaxies, with the latter ones being systematically larger than the MW. These values of global properties suggest that SNA are more likely to thrive in galaxies that meet well defined physical conditions.

\begin{table}
\caption{{Median global and local properties for the MW and the SN}, respectively, compared to the SNA hosts and regions values retrieved in this letter.\label{tab:sna-chars}}
\centering
\begin{tabular}{lrr}
\hline
\noalign{\smallskip}
{} &              {MW/SN} &             {CALIFA/SNA} \\
\hline
\noalign{\smallskip}
{}                                              &  \multicolumn{2}{c}{Global}       \\
\cline{2-3}
\noalign{\smallskip}
Type                                            &     S(B)b~--~bc &      S(B)a~--~c \\
$\log{M_\star/\textrm{M}_\odot}$                &  $10.70\pm0.09$ &  $10.95\pm0.26$ \\
$\log{\psi_\star}$                              &   $0.22\pm0.05$ &   $0.41\pm0.43$ \\
$r_\mathrm{eff}$                                &   $4.20\pm0.40$ &   $5.18\pm3.32$ \\
\hline
\noalign{\smallskip}
{}                                              &  \multicolumn{2}{c}{Local}        \\
\cline{2-3}
\noalign{\smallskip}
$\left<\log{t/\textrm{yr}}\right>_{L_\star}$    &   $9.38\pm0.09$ &   $9.10\pm0.17$ \\
$\left<\log{t/\textrm{yr}}\right>_{M_\star}$    &   $9.70\pm0.22$ &   $9.76\pm0.14$ \\
$\left<[Z/\textrm{Z}_{\odot}]\right>_{L_\star}$ &  $-0.08\pm0.18$ &  $-0.20\pm0.09$ \\
$\left<[Z/\textrm{Z}_{\odot}]\right>_{M_\star}$ &  $-0.06\pm0.20$ &  $-0.17\pm0.14$ \\
$A_V$                                           &   $0.19\pm0.02$ &   $0.15\pm0.25$ \\
$\log{M_\star/L_\star}$                         &   $0.36\pm0.04$ &   $0.60\pm0.12$ \\
$\log{\Sigma_{M_\star}}$                        &   $1.76\pm0.07$ &   $1.85\pm0.27$ \\
$\log{\Sigma_{\psi_\star}}$                     &  $-8.24\pm0.09$ &  $-8.44\pm0.50$ \\
$r_\mathrm{SNA}/r_\mathrm{eff}$                 &   $1.95\pm0.19$ &   $2.05\pm1.61$ \\
\noalign{\smallskip}
\hline
\end{tabular}
\end{table}

Even though the global properties explored above have restricted the type of galaxies in which SNA are likely to be found, local properties have the potential to unveil the environmental physics of such regions. The local conditions of the SN have been widely studied in recent years and we have revealed a clear picture of our vicinity through a wide variety of stellar tracers \citep[e.~g.,][to name a few]{BonattoBica2011, Sollima2019, SahinBilir2020, WhittenPlacco2021, GontcharovMosenkov2021}. The SN is known to be in an expanding local under-density called the local bubble ($<200\,$pc) with little to none star formation within \citep[e.~g.,][]{ZuckerGoodman2022}, low dust extinction \citep[e.~g.,][]{GaiaCollaborationBabusiaux2018}, stellar metallicity around solar value \citep[e.~g.,][]{HaydenBland-Hawthorn2020} and an average (mass-weighted) stellar age $\left<\log{t/\textrm{yr}}\right>_{M_\star}\sim9.5$ \citep[e.~g.,][]{ReidTurner2007}. Outside this vicinity ($<1\,$kpc) the solar neighborhood star formation takes place at a rate per square parsec of $\log{\Sigma_{\psi_\star}}\sim-8.24\pm0.09\,$M$_\sun\,$yr$^{-1}\,$pc$^{-2}$;
this yields a light dominant stellar component with age $\left<\log{t/\textrm{yr}}\right>_{L_\star}\lesssim9.38\pm0.09\,$ and metallicity $\left<[Z/\textrm{Z}_\sun]\right>_{L_\star}\lesssim\,-0.08\pm0.18$ {(Fig.~\ref{fig:sna-pars}c)}; with a surface mass density $\log{\Sigma_{M_\star}}\sim1.76\pm0.07\,$M$_\sun\,$pc$^{-2}$ and mass-to-light ratio in the $V$-band $\log{M_\star/L_\star}\sim0.36\pm0.04\,$M$_\sun\,$L$_\sun^{-1}$ \citep{FlynnHolmberg2006} and; affected by a typical visual dust extinction $A_V\sim0.19\pm0.02\,$mag. 


%
%
In most of the cases we find an astonishing agreement in their local properties, despite of the different methodologies adopted to derive the SNA values (stellar population synthesis of unresolved stars), and the SN ones \citep[exploring the properties of resolved stellar populations; see e.~g.,][]{ReidTurner2007, HaydenBland-Hawthorn2020, GontcharovMosenkov2021, LacerdaSanchez2022}.
Of the compared properties we found difference above $2\sigma$ for only the $M_\star/L_\star$ ratio, a parameter that is particularly difficult to derive for resolved stellar populations \citep[e.~g.,][]{FlynnHolmberg2006}.

As indicated before we find that most of the SNA regions are located following a ring-like distribution around the galactic nucleus (e.g., Fig.~\ref{fig:sna-showcase}).  This ring seems to be located at different galactocentric distances for each galaxy, as suggested by \citet{Prantzos2008} for the MW. We noted above that this distance seemingly scales with the size of the galaxy. In order to check for this possibility, we present in {Fig.~\ref{fig:sna-pars}d} the distribution of the galactocentric distances for the SNAs normalized by the effective radius of the host galaxy.
The fact that the SNA distances distribution can be characterized by a typical value is evidence that such regions tend to thrive around a typical scaled radii value, albeit with a significant scatter, $\sigma_{r_\textrm{SNA}}\sim1.79\,r_\textrm{eff}$.
According to our estimates of the MW effective radius, the SN is located at a scaled galactocentric distance of $r_\textrm{SN}\sim1.95\pm0.19\,r_\textrm{eff}$, assuming a Sun absolute galactocentric distance $\sim8.2\pm0.1\,$kpc \citepalias{Bland-HawthornGerhard2016}. Interestingly, the {median} distance of SNAs and that of the Sun are $\sim2$, and are apart by $\sim0.1\,r_\textrm{eff}$. 

\section{Discussion and conclusions}\label{sec:conclusions}

Most of the efforts to characterize the GHZ in the MW are based on simulations of galaxy formation and chemical evolution.
\citet{LineweaverFenner2004} simulated the physical conditions to produce a MW-like galaxy.
They predicted that the Sun lies well within a locus of GHZ, which encompasses a stellar population with an age $\lesssim4\,$~--~$8\,$Gyr ($\log{t/\textrm{yr}}\lesssim9.6$~--~$9.9$) and galactocentric distances within $\sim5$~--~$11\,$kpc ($1.19$~--~$2.62\,r_\textrm{eff}$). \citet{SpitoniMatteucci2014, SpitoniGioannini2017} developed chemical evolution simulations of the MW and M$31$ to predict the probability of finding terrestrial planets with the conditions to harbor life.
In agreement with \citet{LineweaverFenner2004}, they also found that earth-like planets are more likely in the present than in the past, peaking near $8\,$kpc in galactocentric distance. Interestingly, \citet{SpitoniMatteucci2014} finds for M$31$ that earth-like planets are prone to be found at systematically larger galactocentric radii $\sim16\,$kpc. Similarly, \citet{CarigiGarcia-Rojas2013} found $\sim12$~--~$14\,$kpc for the same galaxy. {Assuming a $r_\textrm{eff}\sim9\,$kpc for M$31$, these radii are within $1\sigma$ of our distribution in Fig.~\ref{fig:sna-pars}d}. These results are {also} consistent with earlier characterizations of the GHZ in our galaxy made by \citet{LineweaverFenner2004, Prantzos2008}. They argued that, as the Galactic disk evolves inside-out, the ring-shaped GHZ spreads toward inner and outer regions of the MW. The picture drawn from these simulations is consistent with our findings that the SNAs in galaxies other than the MW follow a ring-like structure, which radii depends on the size of the galaxy. Consequently, the GHZ (as sampled by our SN definition) should be related to the time-scale at which these galaxies are currently evolving, as gathered from their current total stellar mass and metallicity, SFR, etc.

In summary, the newly introduced methodology has allowed us to find SNAs using a non-parametric method. Most of the local properties of those regions and the global ones of their hosts agree with the those of the SN and the MW. Furthermore, SNAs are found in a ring at a galactoentric distance that scales with the size of the galaxy. This result agrees with both our current understanding of the evolution of the stellar populations in galaxies \citep[e.~g.,][]{Sanchez2020} and simulations predicting the optical location of both the GHZ and the presence of terrestrial planets. A study of the structural evolution of these potential GHZs at higher redshifts is in order.



\begin{acknowledgements}

We thank the support by the PAPIIT-DGAPA grant AG100622. AMN thanks the support from the DGAPA-UNAM postdoctoral fellowship. LC thanks support from (PAPIIT-DGAPA, UNAM), grant IN-103820. JBB acknowledges support from the grant IA-101522 (PAPIIT-DGAPA, UNAM) and funding from the CONACYT grant CF19-39578. We thank the anonymous referee for his/her helpful comments.

This study makes uses of the data provided by the Calar Alto Legacy Integral Field Area (CALIFA) survey (\url{http://califa.caha.es}). CALIFA is the first legacy survey being performed at Calar Alto. The CALIFA collaboration would like to thank the IAA-CSIC and MPIA-MPG as major partners of the observatory, and CAHA itself, for the unique access to telescope time and support in manpower and infrastructures. The CALIFA collaboration thanks also the CAHA staff for the dedication to this project.

Funding for the Sloan Digital Sky 
Survey IV has been provided by the 
Alfred P. Sloan Foundation, the U.S. 
Department of Energy Office of 
Science, and the Participating 
Institutions. 

SDSS-IV acknowledges support and 
resources from the Center for High 
Performance Computing  at the 
University of Utah. The SDSS 
website is www.sdss.org.

SDSS-IV is managed by the 
Astrophysical Research Consortium 
for the Participating Institutions 
of the SDSS Collaboration including 
the Brazilian Participation Group, 
the Carnegie Institution for Science, 
Carnegie Mellon University, Center for 
Astrophysics | Harvard \& 
Smithsonian, the Chilean Participation 
Group, the French Participation Group, 
Instituto de Astrof\'isica de 
Canarias, The Johns Hopkins 
University, Kavli Institute for the 
Physics and Mathematics of the 
Universe (IPMU) / University of 
Tokyo, the Korean Participation Group, 
Lawrence Berkeley National Laboratory, 
Leibniz Institut f\"ur Astrophysik 
Potsdam (AIP),  Max-Planck-Institut 
f\"ur Astronomie (MPIA Heidelberg), 
Max-Planck-Institut f\"ur 
Astrophysik (MPA Garching), 
Max-Planck-Institut f\"ur 
Extraterrestrische Physik (MPE), 
National Astronomical Observatories of 
China, New Mexico State University, 
New York University, University of 
Notre Dame, Observat\'ario 
Nacional / MCTI, The Ohio State 
University, Pennsylvania State 
University, Shanghai 
Astronomical Observatory, United 
Kingdom Participation Group, 
Universidad Nacional Aut\'onoma 
de M\'exico, University of Arizona, 
University of Colorado Boulder, 
University of Oxford, University of 
Portsmouth, University of Utah, 
University of Virginia, University 
of Washington, University of 
Wisconsin, Vanderbilt University, 
and Yale University.

\end{acknowledgements}

\bibliographystyle{aa}
\bibliography{finding-sna, external}

\end{document}